\documentclass[12pt]{article}
\usepackage[T1]{fontenc}
\usepackage[latin1]{inputenc}

\makeatletter

\providecommand{\LyX}{L\kern-.1667em\lower.25em\hbox{Y}\kern-.125emX\@}
\newcommand{\be}{\begin{equation}}
\newcommand{\ee}{\end{equation}}
\newcommand{\bea}{\begin{eqnarray}}
\newcommand{\eea}{\end{eqnarray}}

\makeatother

\begin{document}

\thispagestyle{empty}
\begin{flushright} May 2001 \\
IC/2001/34
\end{flushright}
\vspace{0.3in}

\begin{center}
{\Large \bf New Dimensions New Hopes \\}
\vspace{0.7in}
{\bf Utpal Sarkar \footnote{e-mail: Utpal.Sarkar@CERN.CH}\\}
\vspace{0.3in}
{\sl \small Physics Department, Visva Bharati University,
Santiniketan 731235, India\\}
\vskip .1in
{\sl \small Abdus Salam Insternational Centre for Theoretical Physics,
34100 Trieste, Italy}
\vspace{0.8in}
\end{center}

\begin{abstract}\

We live in a four dimensional world. But the idea of unification of fundamental
interactions lead us to higher dimensional theories. Recently a new theory with
extra dimensions has emerged, where only gravity propagates in 
the extra dimension
and all other interactions are confined in only four dimensions. This theory
gives us many new hopes. In earlier theories unification of strong, weak
and the electromagnetic forces was possible at around 
\( 10^{16} \) GeV in a grand unified theory (GUT) and it 
could get unified with gravity at around the Planck scale of
\( 10^{19} \) GeV. With this new idea it is possible to bring
down all unification scales within the reach of the next generation
accelerators, i.e., around $10^4$ GeV. 

\end{abstract}
\newpage
\baselineskip 16pt

\section{Introduction}

In particle physics we try to find out what are the fundamental particles and
how they interact. This is motivated from the belief that there must be some
fundamental law that governs everything. The diverse phenomenon we see all around
us must somehow be connected and there must be some fundamental principle that
dictates the behaviour of everything. This inherent idea of unification is the
foundation stone of particle physics.

Initially this started as a philosophical question. Thales said everything is
made of water. Then Aristotle told everything is made of water, earth, air and
fire. A similar notion was prevailing in Indian philosophy which said everything
is made of water, earth, air, fire and mind (to distinguish between living and
non-living beings). Around the same time Democritus believed that everything
is made of atomos, the smallest particles which retain the behaviour of the
particle itself. Finally almost after two millennium we come back to this idea
when experiments tell us that this is how nature works. When this idea comes
as a result of experimental findings, it comes out of philosophical domain to
the realm of scientific research.

We did not stop when atoms were found to be the smallest particles. We discovered
atoms are made of protons, neutrons and electrons. We continued to look for
substructure and find that neutrons and protons are made of quarks. Our present
knowledge tells us that the most fundamental particles are quarks and leptons.
They have some intrinsic half integer spin and hence they are the fermions.
There are also spin one particles, which mediate the interactions between these
particles. They are called the gauge bosons, since their nature is dictated
by the underlying gauge theory.

For example, all charged particles undergo electromagnetic interaction, which
is a \( U(1) \) gauge theory. This theory has only one generator, the electric
charge \( Q \). When a charged particle is placed at any point, it creates
a field around it. This information is carried by the generator of the \( U(1) \)
symmetry group, which is the photon. The photon is then the gauge boson 
corresponding to this \( U(1) \) gauge symmetry.

Now consider a non-abelian symmetry group \( SU(3) \). The symmetry group has
a fundamental representation of three elements. There are eight generators of
the group, out of which two are diagonal. These two diagonal generators determine
the color charge of the group and the different elements of the group are connected
by the generators of the group. The quarks undergo strong interactions,
which is a \( SU(3) \) gauge interaction \cite{1}. The quarks are confined inside a
proton and a neutron by the strong interaction. This means there are three 
fundamental quarks, whose interactions are mediated by eight gauge bosons 
corresponding to the generators of the group. These eight gauge bosons are 
called the gluons. The $SU(3)$ color symmetry will ensure that all the
interactions are governed by one coupling constant, the $SU(3)$ gauge
coupling constant. 

All these gauge bosons we mentioned are massless, since these gauge symmetries
are exact. There exists another interaction,
which give rise to beta decay, called the weak interaction. In a beta decay,
a neutron inside a nucleus decay into 
\[ n\rightarrow p+e^{-}+ \bar \nu .\]
 This process can be thought of as a neutron going into a proton releasing a
gauge boson (which has to be charged) which then decays into an electron and
a neutrino. These two charged gauge bosons \( W^{\pm } \) are massive and
has a mass of around 100 GeV. There is another neutral gauge boson, which also
has similar mass. They mediate the weak interaction \cite{2}. 

In the Standard Model the electromagnetic and the weak interactions have been
unified \cite{3}. It is a \( SU(2)_{L}\times U(1)_{Y} \) gauge theory, spontaneously
broken to a \( U(1)_{Q} \). The left handed leptons, an electron and a neutrino,
form a doublet of the group \( SU(2)_{L} \). Similarly a left handed up quark
and a down quark also form a doublet. The gauge bosons connect the two states.
So, if a left handed electron is transformed into a neutrino, a \( W^{-} \)
gauge boson will be released, or a \( W^{+} \) gauge boson will interact with
a down quark to transform it into an up quark. These two groups \( SU(2)_{L} \)
and the \( U(1)_{Y} \) are spontaneously broken by Higgs mechanism and only
a combination of the two neutral gauge bosons remains unbroken. This unbroken
\( U(1)_{Q} \) group generates the electromagnetic interaction. The spontaneous
symmetry breaking give masses to the three gauge bosons. This symmetry breaking
requires another scalar spin 0 particle, called the Higgs boson. This Higgs
boson gets a vacuum expectation value to break the symmetry. The two gauge
groups $SU(2)_L$ and $U(1)_Y$ have two gauge coupling constants. One combination
of them becomes the electric charge (the gauge coupling constant of the unbroken
$U(1)_Q$ symmetry generating the electromagnetic interaction at low energy.

To understand this symmetry breaking, consider a simple example of a boy with
five girlfriends. He will have equal probability to be found with any one of
these five girl friends. So, the boy will have a fivefold symmetry. Now, if
he gets married to any one of them, this symmetry will be broken and the boy
will be trapped in the infinite potential well of his wife.
The fivefold symmetry in his movement will be lost and his movement will be
restricted only around his wife, which is his vacuum expectation value.

In the standard model, there are three coupling constants, corresponding to
the three gauge groups \( SU(3)_{c} \), \( SU(2)_{L} \) and \( U(1)_{Y} \).
All the
gauge bosons of any group couple to fermions belonging to its fundamental 
representations
with the same coupling constant. These coupling constants change with energy.
It was then found that at very high energies (about \( 10^{16} \) GeV), these
coupling constants approach a single point. At this energy it is possible to
have a single gauge group with only one gauge coupling constant, which contains
all the low energy subgroups. This is the so-called grand unified theory (GUT). All
the three interactions, the strong, the weak and the electromagnetic interactions
come out of this single grand unified interaction as its low energy manifestations.
At the grand unification scale of about \( 10^{16} \) GeV, there are new Higgs
scalars, which give rise to a symmetry breaking, whereby the grand unified group
\( {\mathcal{G}} \) breaks down to its low energy subgroups 
\[
{\mathcal{G}}\longrightarrow SU(3)_{c}\times SU(2)_{L}\times U(1)_{Y}\]
 All the fermions, the quarks and the leptons, now belong to some representations
of the grand unified group \cite{4}.

In a grand unified theory there are now new gauge bosons which can relate a
quark to a lepton, giving rise to proton decay. However, since the scale of
unification is very large, these processes are suppressed and 
the lifetime of proton comes out to be greater
than \( 10^{32} \) years. But the baryon number violation due to this grand
unification has another very good implications. This can explain why there are
more baryons compared to antibaryons in the universe we see around us. This
generation of baryon asymmetry of the universe came out as a bonus to this theory
\cite{5}.

This unified picture of all gauge interactions has very strong theoretical 
motivation, but experimentally it could not be verified. Due to the limitations 
of the accelerator
energies, direct verification is completely out of question. Moreover, there
is one serious problem with this theory. The existence of such large scale and
the Higgs scalars with such large vacuum expectation values, tends to make the
electroweak Higgs scalars superheavy. This two very widely different scales
then pose the problem of gauge hierarchy. This problem states, what prevents
the Higgs scalars required for the electroweak symmetry breaking from picking
up a mass of the order of the very large scale, say the the scale of grand
unification. 

A solution to this gauge hierarchy problem \cite{6} then gave birth to a new theory,
called supersymmetry \cite{7}. This is a symmetry between a fermion and a boson which means
that corresponding to every fermion, there is a boson with equal mass and 
corresponding
to every boson there is a fermion with same mass. However, such superparticles
have not been observed in nature so far. So, this supersymmetry has to be broken
at some scale close to the electroweak symmetry breaking scale of around 100
GeV. This will mean that all the superparticles, \textit{i.e.}, the superpartners
corresponding to the particles we see around us, gets a mass at the supersymmetry
breaking scale. So, although we have not seen these particles so far, in the
next generation accelerators we should see these particles.

The supersymmetric grand unified theory can then be the consistent unified theory
of all gauge interactions \cite{8}, which could be tested in the next generation 
accelerators.
Although we cannot reach the unification scale experimentally, the existence
of the superparticles and other low energy signatures can give us enough evidence
for this theory. But this theory does not include gravity at any stage. Then
the question comes how to unify gauge interactions with gravity.

\section{Kaluza-Klein Theory}

The first approach to unify gravity with any gauge interaction was due to Kaluza
and Klein \cite{9}. They tried to explain both gravitational interaction and the 
electromagnetic
interaction from one theory. They work in a five dimensional space-time, where
one of the space dimensions are compact. For example, the boundary of a circle
is a one dimensional compact space. If we identify the two end points of a line
with each other, it can be identified with a circle. Similarly, any space with
all its end points identified to each other becomes a compact space, since in
no direction one can reach the point at infinity. When any such compact space
dimension is viewed from a distance it appears as a point. Consider the boundary
of a circle. If we look at it from larger and larger distances, it will start
appearing smaller and smaller. Finally it will appear as a point. This is the
basic idea of Kaluza-Klein theory. Although there are four space dimensions,
one of the space dimensions is compact with a very small radius. As a result,
in all experiments we could see effects of only four dimensions. Since there
are now four space dimensions and one time dimension, we have to write the theory
of gravity in five dimensions. The mediator of gravity, the graviton, expressed
by the space-time metric in a linearized theory of gravity, will now contain
the four dimensional graviton and also the photon, which mediates the electromagnetic
interaction. This idea can then be extended to include all gauge interactions
in the higher dimensional metric.

The foundation stone for the general theory of relativity is the equivalence
principle \cite{10}, 
which assumes equivalence between the inertial mass with the gravitational
mass. This implies the weak equivalence principle, that effects of gravitation
can be transformed away locally by using suitably accelerated frames of reference.
This can then be generalized to strong equivalence principle, which allows us
to study gravitational interaction by studying only the geometry of the space-time. 

Let us consider an example of a flat space-time in the absence of any
matter. Compare this with a wire mesh, stretched from its end points. If you
place any object the space-time will no longer remain flat. It is like placing
an object on the wire mesh (see figure 1). Now one can study the effect of
this object in this space-time only by considering the curvature of the 
space-time. For example, if you place another small ball on the wire mesh,
it will roll down towards the big ball. In the language of gravity, we may 
say that if the space has a curvature like this (as created by the big ball), 
then any other smaller 
object will role down along the curvature, or in simple language two objects
always attract each other under the influence of gravity. 

\begin{figure}[htb]
\mbox{}
\vskip 2in\relax\noindent\hskip .2in\relax
\includegraphics{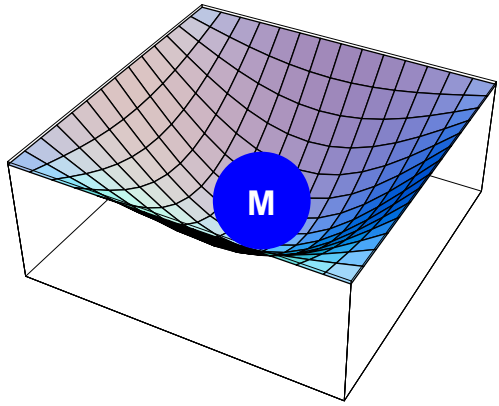}
\caption{Curvature of space in the presence of a massive object M.}
\label{curvMfig}
\end{figure}

To understand the geometry of space-time, consider the distance between two
space-time points in any flat inertial space, 
\begin{equation} 
ds^2 = dt^2 - dx^2 -dy^2 -dz^2. 
\end{equation} 
But
if these two points are not connected by straight line, the distance can be given
by a more general form, 
\begin{equation} 
ds^2 = g_{\mu \nu} dx^\mu dx^\nu 
\end{equation} 
where
sum over repeated index is implied. The indices \( \mu, \nu =0,1,2,3 \) run
over four space-time coordinates. The coefficient \( g_{\mu \nu } \) is a
function of the space-time coordinate \( x^{\mu }. \) This is called the metric
and this specifies the geometry of the space-time. To study the geometry of
any space-time for an understanding of the theory of gravity in this space,
it is enough to study the metric \( g_{\mu \nu }. \) 

Consider a line element in a five dimensions, which can be expressed in terms
of a five-dimensional metric \( g_{MN} \) (with \( M,N=0,1,2,3,4 \) ) as,
\begin{equation} ds^2 = g_{MN} dx^M dx^N. \end{equation} We then make an ansatz
to express this five dimensional line element in terms of four dimensional metric
and a four vector \( A_{\mu } \) as 
\begin{equation} 
ds^2 = g_{\mu \nu} dx^\mu dx^\nu -(dy-\kappa A_\mu(x) dx^\nu)^2 . 
\end{equation} 
This
then becomes invariant under a gauge transformation, defined as a coordinate
transformation, 
\begin{eqnarray} x^\mu & \to & x^\mu \nonumber \\ 
y & \to & y + \Lambda(x) \nonumber \\ 
g_{\mu \nu} & \to & g_{\mu \nu} \nonumber \\ 
A_\mu(x) & \to & A_\mu(x) + {1 \over \kappa} \partial_\mu \Lambda . \label{gauge} 
\end{eqnarray} The
gauge covariant derivative is now defined as, 
\begin{equation} 
D_\mu = \partial_\mu - i {\kappa \over r} A_\mu 
\end{equation} 
where
the electric charge is given by, 
\begin{equation} 
Q = {\kappa \over r}, 
\end{equation} 
where
\( r \) is the radius or size of the compact fifth dimension. This clearly
indicates that the coordinate transformation in the fifth dimension can appear
as gauge transformation in four dimensions. 

Above ansatz is valid only when the extra dimension is compact, i.e., the
end points can be identified by the boundary condition 
\begin{equation} y = y+ 2 \pi r. \end{equation} 
A
scalar field satisfying this boundary condition 
\begin{equation} \phi(x,y) = \phi(x,y+ 2 \pi r) \end{equation} 
can then be expanded in a Fourier series 
\begin{equation} 
\phi(x,y) = \sum_{n=-\infty}^\infty \phi_n(x) e^{i n y/r} . 
\end{equation} Single-valuedness
of the exponent under the boundary condition of \( y \) will then imply that
the transverse momentum of any state \( \phi _{n} \) will be \( p\sim O(n/r) \).
So, in four dimensions we shall see all these excited states with momentum
\( \sim O(n/r). \)
Since we want to unifiy the electromagnetic interactions with gravity, the natural
radius of compactification will be the Planck length 
\begin{equation} r={1 \over M_P}, \end{equation} 
where
the Planck mass \( M_{P}\sim 10^{18} \) GeV. Thus only the zero modes \( (n=0) \)
will be observable at our present energy and all the excited states will have
masses of the order of the Planck scale. Only when we reach that energy we can
see all these excited states. At the Planck scale we shall also be able to 
resolve the extra dimension and at even higher energies this fifth dimension
will appear to be similar to the other space dimensions. 

Similar to the scalar field the five dimensional metric $g_{MN}(x,y)$ can also
be expanded in a power series,
\begin{equation}
g_{MN}(x,y) = \sum_{n=-\infty}^\infty g_{MN}^{(n)}(x) e^{i n y/r} . 
\end{equation} 
The $n=0$ mode can then be parametrized as,
\begin{equation}
g_{MN}^{(0)} = \phi^{-1/3} \pmatrix{ g_{\mu \nu} + \phi A_\mu A_\nu & 
\phi A_\mu \cr \phi A_\nu & \phi }
\end{equation}
where the field $\phi$ appears as a scaling parameter in the fifth dimension
and is called the dilaton field. In the power series expansion, all the
higher modes will have mass of the order of multiples of Planck mass. Only
the lowest state will remain massless. 
The $n=0$ massless mode of the gauge field $A_\mu$
can now mediate the electromagnetic interaction. 

We can then start with the Einstein action in five dimensions with the
five dimensional metric. Make the ansatz for the metric and assume a
compact fifth dimension and then expand the metric in terms of its
four dimensional zero modes. In four dimension we then get the usual
Maxwell action for the electromagnetic interaction and the Einstein
action in four dimension explaining the gravitational interaction.
Thus starting from
a five dimensional gravitational interaction, given by metric in the five
dimensional space-time, we get the four dimensional metric (giving us the
gravitational interaction) and the four vector (giving us the electromagnetic
interaction). 

\section{Developments in higher dimensional theories}

The Kaluza-Klein theory was then extended to include all the gauge 
interactions. The electromagnetic interaction could be accomodated
with only one extra dimension. But to accomodate the strong, weak
and electromagnetic interactions, i.e., the $SU(3)_c \times SU(2)_L
\times U(1)_Y$ gauge theory, we need at least 7 extra dimensions. 
So, many attempts were made to study a eleven dimensional theory,
which can then give rise to all these interactions in our four
dimensional world. This 11 dimensional unified theory has another
importance. 

In an ordinary supersymmetric theory, we assume there is only one
supersymmetric charge. So, every boson is related by this charge to 
a fermion and any fermion is related to a boson by the same charge. 
This is called a $N=1$ supersymmetry. There could be higher 
supersymmetric theories as well. $N=2$ supersymmetry will have two
charges, $Q_1$ and $Q_2$. If, $Q_1$ takes a scalar to a fermion,
$Q_2$ will take the fermion to a vector. Similarly, $Q_2$ takes the
scalar to another fermion, which gets related to the same vector 
by the action of the charge $Q_1$. Thus in the smallest multiplet 
there is one scalar, two fermions and one vector. The next higher
theory is the $N=4$ supersymemtry, in which the lowest multiplet 
contains a spin--3/2 particle, while the $N=8$ supersymmetry contains
a spin--2 particle in the lowest multiplet. Since there are no 
consistent field theory for particles with spin higher than 2,
$N=8$ supersymmetry is the largest supersymemtry we can consistently
construct. It was then shown that $N=8$ supersymmetric theory in
four dimension has a correspondence with the 11-dimensional 
$N=1$ supergravity theory. Any higher dimensional supersymmetric 
theory would then be inconsistent, since they will have 
correspondence with higher than $N=8$ supersymmetric theories. 
From this point of view, 11-dimensional one is the highest dimensional
theory which could be constructed consistently. 

The 11-dimensional supergravity theory was studied extensively. 
But it was realised that it is not possible to get all the gauge
interactions and the required fermion contents of the standard model
from this theory. Then there was attempt to consider 11-dimensional
theories with gauge groups. Of course, the main motivation of 
obtaining all gauge interactions and gravity from one Einstein
action at 11-dimensions would be lost, but still this became an
important study for some time. In this construction the main 
problem was due to a new inconsitency, the anomaly. 

In any gauge theory there are some triangle loop diagrams 
(with fermions in the loop and gauge bosons at the vertices) which are
diverging. It depends on the fermion contents of the theory.
The symmetry of the classical lagrangian is then broken by
quantum effects, if these anomaly diagrams are non-vanishing
in any theory. Thus for theoretical consistency one needs to
make any gauge theory anomaly free. In higher dimensions, one
has to take care of gauge anomalies as well as the gravitational
anomalies. In any 11-dimensional supersymmetric theory, if there is a 
gauge group, the gauge fermions would contribute to the 
gauge and gravitational anomalies. If these anomalies do not
cancel, the theory would be inconsistent. Thus the anomaly problem 
could not be solved in 11-dimensional theories. 

This problem was solved in a 10-dimensional superstring theory.
If one considers an extension of the
particles along one internal extra dimension, so that the 
particles appear as strings with certain boundary conditions
(to take care of the problem of causality), then one can construct
a consistent field theory for these strings. A very nice feature
of such superstring theories is that, in 10-dimensions the
gauge and gravitational anomalies cancel for the $E_8 \times E_8$
group and the $SO(32)$ group \cite{11}. It was then found that when the 
extra six dimensional space is compactified, the four dimensional
world contains all the required fermions and the standard model
gauge groups. Supersymmetry could remain unbroken till the 
electroweak scale to take care of the gauge hierarchy problem.
This then appears to be the unified theory of all known 
interactions \cite{12}.

It is not easy to obtain all the low energy features starting
from the superstring theory at 10-dimensions. But the superstring
theory has established itlself as the most consistent theory of
quantum gravity. There are several types of superstring theories,
but now some duality conjectures have related all these theories
to one consistent theory in 11-dimensions. So, although these 
theories now appear to be far from any experiments, it is now
established that these theories has the prospect of becoming
the theory of everything. The scale at which this theory is
operational is close to the Planck scale. This makes it 
experimentally non-viable for a very very long time, or probably
at any time. 

\section{New theories with extra dimensions}

While all these interesting ideas of unification were predicting 
very high energy scale, which cannot be reached by any Laboratory
experiments, one very interesting theory emerged with all the
new hopes \cite{13,14,15}. This theory predicts all new physics within the
range of our next generation accelerators. If any signatures of 
this theory is observed in nature, it will mean that we can see
signals of superstring theory, black holes, grand unified theories,
all in the next generation accelerators \cite{16}. 

It started from our understanding of membranes. In a string theory,
we assume that the particles we see around us is actually like
strings. Since the entire string propagates with time, we have to
apply boundary conditions to the end points for consistency. This
lead us to either open or closed strings, which has different 
boundary conditions.
When this theory was extended to a membrane, one has to apply 
boundary conditions to its boundary surfaces. This can then be
extended to higher n-dimensional branes. In general, branes are
static classical solutions in string theories. A p-brane denotes 
a static configuration which extends along p-spatial directions
and is localized in all other directions. Strings are equivalent 
to 1-branes, membranes are 2-branes and particles are 0-branes. A
p-brane is described by a $(p+1)-$dimensional gauge field theory. 

In the new class of theories with large extra dimensions \cite{14,15}, one 
uses this idea of p-branes and assume that the standard model
particles are confined along the three spatial dimensional walls of a 
higher dimensional theory. Only gravity propogates along the
bulk or all the dimensions. Since such solutions
are invariant under translation in the transverse directions, the
standard model particles cannot feel the extra dimensions. Since
the standard model particles are stuck to the three dimensional
walls of a higher dimensional theory, they are explained by four 
dimensional gauge field theory. Only the $4+n$--dimensional gravitons
are free to propagate in all the directions along the bulk. The
thickness ($M^{-1}$) of the boundary walls sets the scale ($M$) 
of the theory to be around a few TeV. Since only gravity
propagates along the bulk, any experiments with the standard
model particles cannot detect the existence of these extra 
dimensions. This allows these extra dimensions to be very
large, of the order of a few mm. This will then cause deviation
from the Newton's law at a distance of a few mm. 

The main feature of the theory with large extra dimensions is
that it predicts the scale at which gravity can get unified with other
interactions to be about a few TeV and not the Planck scale. In
other words, since gravity propagates mostly along the extra dimensions,
in our 3-brane it has a very little probability, which suppress its
coupling to ordinary particles. But the gravity couplings in the
bulk is not small. It is assumed that such confinement takes
place at some high energy. So, at higher energies, all the dimensions
becomes transparent to the standard model particles and then 
gravity becomes as strong as other interactions. So, the
scale at which gravity interacts strongly in our world 
becomes the fundamental scale of this theory, which
is of the order of a few TeV. 

Consider a $4+n$ dimensional space, where the extra $n$ dimensions
are compact with radius $R$. The gravitational potential in the
four dimensional world between two test particles with masses $m_1$ and $m_2$
separated by a distance $r \ll R$ is given in terms of the Planck
scale $M$ (through its relation to the Newton's constant) as
\begin{equation}
V(r) \sim {m_1 m_2 \over M^{n+2} } {1 \over r^{n+1} } .
\end{equation}
When these test particles are separated by large distance $r \gg R$,
the gravitational flux cannot penetrate the extra dimensions and hence
the potential would be
\begin{equation}
V(r) \sim {m_1 m_2 \over M^{n+2} R^n } {1 \over r } .
\end{equation}
In our four-dimensional world we shall then get the effective potential
\begin{equation}
V(r) \sim {m_1 m_2 \over M_{Pl}^{2} } {1 \over r } ,
\end{equation}
which relates the true Planck scale $M$ in the higher-dimensional theories and
the effective Planck scale $M_{Pl}$ of our  4-dimensional space through
\begin{equation}
M_{Pl}^2 = M^{2 + n} R^n .
\end{equation}
Using the observed Planck scale in 4 dimensions, we can get an estimate
of the radius of compactification of the extra dimensions. 
If we now require that M is of the order of TeV, then
$n=1$ is not allowed since it requires a deviation from the Newtonian
gravity at a distance $R \sim 10^{13}$ cm, which is about the distance
scale of the solar system. Since we have tested gravity only down to
a distance of a mm, any value of $n \geq 2$ is allowed. In all earlier theories
M was assumed to be the same as the Planck scale of 4 dimensions, which
made the compactification scale to be of the order of the Planck length.
At present the Newton's law is being tested at this scale and any
deviation would then imply the existence of such large extra dimensions.

The experimentally verifiability of these theories with extra large 
dimensions make them most interesting. In the conventional theories
the gauge couplings constants of the standard model gets unified
at the grand unified scale of about $10^{16}$ GeV. But in theories 
with large extra dimensions this also gets changed. Since the
compactification scale is the fundamental scale of a few TeV,
the Kaluza-Klein excited states corresponding to the standard model
fermions now have mass of multiples of a few TeV. All these new
particles then modify the gauge coupling constant evolution and 
the unification is now achieved at a scale slightly above the 
fundamental scale \cite{13}. The unification scale is now brought down from
about $10^{16}$ GeV to about 30 TeV (see figure 2).
So, we have grand unified theory within the 
reach of the next generation accelerators. In addition to the signals
for the grand unification or quantum gravity, there are other 
signals of the new physics arising due to the existence of the
new dimensions. Since gravitons propagate in the bulk, some of the
energies can be carried away by the gravitons, giving rise to new
signals in colliders with missing transverse energies. 

\begin{figure}[hbt]
\mbox{}
\vskip 4.5in\relax\noindent\hskip .75in\relax
\includegraphics{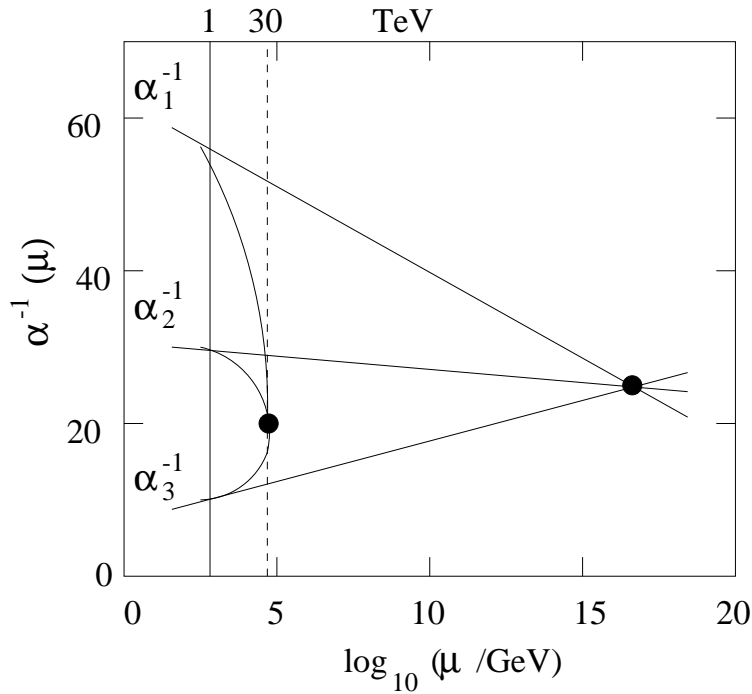}
\vskip -.75in
\caption{Coupling constants evolve along the straight line and
get unified at around $10^{16}$ GeV in ordinary grand unified theories. 
In the presence of extra 
dimensions the Kaluza-Klein excited modes make the coupling constants
evolve very fast above 1 TeV so that they get unified at around 30 TeV.}
\label{exl}
\end{figure}

In these theories there are no large scale and all the very high
energy scales are brought down within our reach. But this has some
drawback. For an explanation of some phenomena we need very high 
scales. For example, the inflation in the early universe, the smallness
of the neutrino mass, the smallness of the axion couplings to solve 
the strong CP problem, etc. These problems now require new solutions.
But such solutions can again lead to new signatures which are testable. 
In the following we discuss one such problem in little details and
its solution \cite{17}.

In the standard model neutrinos are massless. An explanation of the 
tiny neutrino mass, as evidenced by the atmospheric neutrino
anomaly, requires new physics beyond the standard model. Since neutrinos
are charge neutral, they can be ordinary Dirac particles like any
other quarks or leptons, or they can be their own anti-particles. 
In the later case two neutrinos will annihilate each other and
they are called Majorana particles. For a Majorana particle we 
do not require any new right handed neutrino states to give mass.
Only the left handed neutrinos can have a mass term of the form,
\begin{equation}
{\cal L} = m_\nu \overline{\nu^c_L} \nu_L
\end{equation}
Since this violates the $SU(2)_L$ symmetry, this mass is forbidden by
the electroweak interaction. But after the $SU(2)_L$ symmetry is broken 
by the vacuum expectation value of the $SU(2)_L$ doublet Higgs scalar $H$, it
is possible to get this mass term from an effective coupling
\begin{equation}
{\cal L} = {f \over M} \overline{\nu^c_L} \nu_L H H  .
\end{equation}
Since this effective operator has a mass dimension five, it is 
suppressed by some scale $M$. The Majorana mass term breaks lepton 
number and hence the large scale $M$ is the lepton number violating
scale. 

This effective operator can be realised in more than one ways. We
shall discuss only one of these mechanisms, which can be embedded in 
theories with large extra dimensions. We extend the standard model
to include one triplet Higgs scalar, $\xi$, which has the interactions
with the doublet Higgs scalar $H$ and the leptons $\ell$,
\begin{equation}
{\cal L}_\xi = \mu \xi H H + f \xi^\dagger \ell \ell + M^2_\xi \xi^\dagger \xi
\end{equation}
where $\mu$ is some mass parameter of the order of the mass of $\xi$.
The first two terms together violate lepton number explicitly 
at a large scale, since the Higgs doublets $H$ does not carry any 
lepton number, while the left handed leptons $\ell$ carry lepton number 1. 
The vacuum expectation value ($vev$) of the Higgs doublet will then
induce a tiny $vev$ to the triplet Higgs scalar $\xi$, which would
then generate a very small neutrino Majorana mass as required,
\begin{equation}
m_\nu = f \langle \xi \rangle = f \mu {\langle H \rangle^2 \over M_\xi^2} .
\end{equation}
The scale of lepton number violation and hence the mass of 
$\xi$ is required to be of the order of $10^{13}$ GeV to get a 
neutrino mass as required by the atmospheric neutrino problem or
the solar neutrino problem \cite{18}. 

In theories with large extra dimensions, such large scales are not
there. As a result the smallness of the neutrino mass requires new
explanation. In fact, most small numbers cannot be obtained by
large mass scale suppression. However, there are small numbers
in the theory arising from completely different mechanism as we shall 
discuss. In this mechanism of neutrino mass what we need is very
small $\mu$ and $M_\xi$ to be of the order of a few TeV. 

The origin of the small numbers in theories of large extra dimensions 
could be from a distant breaking of some symmetry. If the lepton number
is broken in a brane which is separated from our brane by a
large distance, then the induced lepton number violation in our brane
will be small. In fact, they will be exponentially suppressed by the
distance between the two branes. If the distance between the two 
branes is of the order of radius of compactification, one can make
an estimate of the amount of lepton number violation in our world. 
From this one can calculate the value of $\mu$ and hence the neutrino
masses in this model. 

Since the mass of the triplet Higgs in this model is of the order
of a few TeV, the decays of the triplet Higgs would be observed in
the colliders. The triplet Higgs $\xi$ decays into two same sign
leptons, which is a very clean signal and can be observed in the
Tevatron upgrade or the LHC. If this dilepton signal is observed
and simultaneously a lepton number violation is established in the
neutrinoless double beta decay experiments, it will then imply that
there exists large extra dimensions in nature and neutrinos get mass
from the triplet Higgs scalars. This mechanism has another very
interesting feature. If the dilepton signals are observed in nature, 
the different branching ratios of $\xi$ would give us the couplings
of $\xi$ with the leptons. Since these couplings enter into the
neutrino mass, we shall be able to get all informations about 
neutrino masses from colliders. This mechanism has another 
advantage. If we can observe the dilepton signals and also some
indications for lepton number violation, then it would be an
indirect indication of the existence for the large extra dimensions. 
This signal has the advantage over other signals for large extra
dimensions in the sense that in this case the triplets could be 
light even if the Planck scale is heavier by few orders of
magnitude. This verifiability of this model makes it better
than otherwise equivalent models. 

Similar to the models of large extra dimensions, there is another
theory with small extra dimensions \cite{14}. In this theory the hierarchy
problem is solved with a non-factorizable metric. While the 
Planck scale physics resides in a hidden sector brane, in our
brane all physics is governed by the fundamental scale. The
distance between our brane and the hidden sector brane gives
exponentially suppressed small numbers. Only gravitons propagates
along the bulk, the space connecting the two branes, while the 
standard model particles are confined in our brane. Although
there are many similarity and the basic idea is similar 
between the theory of large extra dimensions and the theory
with warped compactification or small extra dimension, 
technically these two models are widely different. But the
main importance of these ideas comes from the verifiability
of these models since both of them predict a very low Planck
scale and many new signatures in the next generation accelerators. 

The models of extra dimensions have thus brought to us the
physics of very large scale, which were otherwise inaccessible to
our experiments. If this theory happens to be the actual theory of
nature, then the next generation accelerators can find all kinds
of new physics. In other words, the idea of large extra dimensions brought to
us all kinds of new hopes. If we do not see any signals for this
theory with large extra dimensions, then all the new physics will 
remain unknown to us for ages. We may not find out if
there is further unification of forces in nature or not or which is the
actual theory of quantum gravity. Only experiments can tell us 
whether the new extra dimensions are only giving us new hopes or
whether this is the theory of nature. 

{\bf Acknowledgement} I would like to thank Abbas Ali for discussions.

\end{document}